\documentclass[twocolumn,showpacs,showkeys,prd,floatfix,nofootinbib]{revtex4}
\usepackage{ifthen,graphicx}
\usepackage{bm}
\usepackage{amsmath,amssymb}

\newcommand{\ovl}{\overline}

\begin{document}
\title{Mannheim's linear potential in conformal gravity}

\date{\today}

\author{Peter~R.~Phillips}
\affiliation{Department of Physics, Washington University, St.~Louis,
MO 63130 }
\email{prp@wuphys.wustl.edu}

\begin{abstract}
We study the equations of conformal gravity, as given by Mannheim, in the
weak field limit, so that a linear approximation is adequate. Specializing
to static fields with spherical symmetry, we obtain a second-order equation
for one of the metric functions. We obtain the Green function for this
equation, and represent the metric function in the form of integrals over
the source. Near a compact source such as the Sun the solution no longer
has Schwarzschild form. Using Flanagan's method of obtaining a conformally
invariant metric tensor we attempt to get a solution of Schwarzschild type.
We find, however, that the $1/r$ terms disappear altogether. We conclude
that a solution of Mannheim type cannot exist for these field equations. 
\end{abstract}


\maketitle


\section{INTRODUCTION}
\label{sec:intro}

In this paper we will derive solutions, in the weak field limit, of the
field equations of conformal gravity as given by Mannheim (see \cite{mann6},
equation (186); this paper will be referred to as PM from now on):
\begin{equation}
4 \alpha_g W^{\mu \nu} \equiv 4 \alpha_g \left[ W^{\mu \nu}_{(2)}
- \frac{1}{3} W^{\mu \nu}_{(1)} \right] = T^{\mu \nu}
\label{eq:field1}
\end{equation}
Here $W^{\mu \nu}$ is the Weyl tensor, the two separate parts
$W^{\mu \nu}_{(1)}$ and $W^{\mu \nu}_{(2)}$ being defined in PM (107) and
(108). $\alpha_g $ is a dimensionless coupling constant. (We adopt the
notation of Weinberg \cite{wein2}, with units such that $c = \hbar = 1$.) 

The energy-momentum tensor, $T^{\mu \nu}$, is derived from an action
principle involving a scalar field, $S$ (see PM (61)). Appropriate variation of
this action yields $T^{\mu \nu}$ as given in PM (64). In Mannheim's model,
the solutions of the field equations undergo a symmetry breaking transition
(SBT) in the early Universe, with $S$ becoming a constant, $S_0$. Making this
change in PM (64) we obtain
\begin{eqnarray}
T^{\mu \nu} & = & -\frac{1}{6} S^{2}_{0} \left( R^{\mu \nu}
- \frac{1}{2} g^{\mu \nu} R^{\alpha}_{\;\;\alpha} \right)
\nonumber \\
  & & {} - g^{\mu \nu} \lambda S_{0}^{4} + T_{M}^{\mu \nu}
\label{eq:mat1}
\end{eqnarray}
where $T^{\mu \nu}_{M}$ is the matter tensor, containing all the usual
fermion and boson fields.

We break from Mannheim's development at this point. The factor $1/6$ in
(\ref{eq:mat1}) derives from the original, conformally invariant action. A
SBT, however, will not necessarily preserve such relations, and we will
instead write
\begin{eqnarray}
T^{\mu \nu} & = & \frac{1}{8 \pi G_0 } \left( R^{\mu \nu}
- \frac{1}{2} g^{\mu \nu} R^{\alpha}_{\;\;\alpha} \right)
\nonumber \\
  & & {} - g^{\mu \nu} \frac{\lambda_W }{8 \pi G_0 } + T_{M}^{\mu \nu}
\label{eq:mat2}
\end{eqnarray}
so that the field equations can be written
\begin{eqnarray}
W^{\mu \nu} & - & \frac{1}{32 \pi \alpha_g  G_0 } \left( R^{\mu \nu}
- \frac{1}{2} g^{\mu \nu} R^{\alpha}_{\;\;\alpha} \right) 
\nonumber \\
  & & {} + g^{\mu \nu} \frac{\lambda_W }{32 \pi \alpha_g G_0 }
= \frac{1}{4 \alpha_g } T_{M}^{\mu \nu}
\label{eq:field2}
\end{eqnarray}
Mannheim is constrained to get an effective $G$ that is negative. We,
on the other hand, will assume that the SBT results in a positive value
for $G_0$. We can then identify $G_0 $ with the Newton gravitational
constant.

In the rest of this paper we will ignore the term in $\lambda_W $.
Defining $\eta = -1/(32 \pi \alpha_g G_0)$ and $\xi = 1/(4\alpha_g )$,
the field equations become
\begin{eqnarray}
W^{\mu \nu} + \eta \left( R^{\mu \nu}
- \frac{1}{2} g^{\mu \nu} R^{\alpha}_{\;\;\alpha} \right) 
  & = & \xi T_{M}^{\mu \nu}
\label{eq:field3}
\end{eqnarray}
$\xi$ is dimensionless, but $\eta$ has dimension ${\textrm length}^{-2}$, so
its magnitude can be written $|\eta | = 1/r_{0}^{2}$, where $r_0$ divides
lengths into two regimes, in one of which ($r < r_0 $) the Weyl tensor is
dominant, and in the other ($r > r_0 $) the Einstein tensor.

We will call this equation the Weyl-Einstein equation, or ``W-E equation''
for short. In the important special case that $\alpha_g W^{\mu \nu}$ is
negligible, or even identically zero, we regain the usual Einstein
equations, as given, for example, in Weinberg \cite{wein2} (16.2.1).
In the opposite limit, $\eta \rightarrow 0$, we obtain
\begin{eqnarray}
W^{\mu \nu} & = & \xi T_{M}^{\mu \nu}
\label{eq:bach}
\end{eqnarray}
the Bach equation. Some solutions of this have been obtained by Fiedler and
Schimming \cite{fiesch}.

We can take the trace of (\ref{eq:field3}), to get
\begin{equation}
R^{\alpha}_{\;\;\alpha} = 8 \pi G_0 T^{\alpha}_{\;\;\alpha}
\label{eq:trace}
\end{equation}
which is, of course, the same as we would get from the Einstein equations
since $W^{\mu \nu}$ is traceless.

From this Mannheim derives a traceless energy-momentum tensor,
PM (65). We shall not use this, however, because it contains less
information than the original tensor, and must be supplemented by the
trace equation.

No exact solutions of the W-E equation seem to be available,
except for the usual Schwarzschild solution, which satisfies both the
Einstein and the Bach equations independently. In this paper we will ask
whether a different solution also exists. We will not seek an exact
solution, but will restrict ourselves to weak fields, and the linear
approximation to the W-E equation. This should be adequate for studies of
galactic rotation and gravitational lensing, and may give us insight into
what a more complete solution would look like in the region of the
Solar System.

\section{\label{sec:statssym}Static fields with spherical symmetry}

We now specialize further, to static fields with spherical symmetry.
Like Fiedler and Schimming, but apparently independent of them, Mannheim and
Kazanas \cite{mann04} addressed the problem of the solution of the Bach
equation under these conditions. They found that in addition to the usual
$1/r$ term of the Schwarzschild solution there was a term $\gamma r$.
Mannheim has used this linear potential to obtain a fit to the rotation
curves of galaxies; for a recent paper, see \cite{mann8}. However, the
relevant field equation is not the Bach equation, but the W-E equation, for
which a linear potential is not a solution. It is therefore not clear what
use a linear potential can be in such studies, except as an approximation
over a limited range.

The most general form for a static metric with spherical symmetry is given
in Weinberg \cite{wein2}, (8.1.6):
\begin{equation}
d \tau^2 = B(r) dt^2 - A(r) dr^2 - r^2 \left( d\theta^2
+ \sin^2 \theta \,d\phi^2 \right)
\label{eq:sphsymmet}
\end{equation}

For weak fields we write $A(r) = 1 + a(r)$ and $B(r) = 1 + b(r)$,
where $a(r)$ and $b(r)$ are assumed small compared to unity, so that only
terms linear in $a(r)$ and $b(r)$ need be considered.

In the presence of matter with density $\rho(r)$, the trace equation
becomes, with primes denoting differentiation with respect to $r$:
\begin{eqnarray}
\eta R^{\alpha}_{\;\;\alpha} = -\xi T^{\alpha}_{\;\;\alpha}
& = & \xi \rho (r)
\\
-\frac{\eta}{2r^2} \left( 4a - 4r b^{\prime} + 4ra^{\prime}
- 2 r^2 b^{\prime \prime} \right) & = & \xi \rho (r)
\label{eq:trace_2}
\\
2 \left( ra \right)^{\prime} - \left( r^2 b^{\prime} \right)^{\prime}
& = &  -\frac{r^2 \xi \rho (r)}{\eta}
\label{eq:trace_3}
\end{eqnarray}
We assume the density is a smooth function,
so that $a^{\prime}(r)$ and $b^{\prime}(r)$  are both zero at $r=0$.
Then we can integrate out from the origin to $r$ to get
\begin{eqnarray}
2ra - r^2 b^{\prime} & = & -\frac{\xi}{4 \pi \eta}
\int_{0}^{r} 4 \pi u^2 \rho(u)\,du
\nonumber \\
  & = &  -\frac{\xi}{4 \pi \eta}m_{e} (r)
\label{eq:beqn}
\end{eqnarray}
where $m_e (r)$ is the enclosed mass out to $r$.

The $r,r$ component of the W-E equation gives \footnote{For the
geometrical calculations we have used GRTensorII, followed by a Maple script
to extract the linear terms.}
\begin{eqnarray}
-r^3 b^{\prime \prime \prime} - 2r^2 b^{\prime \prime}
- r^2 a^{\prime \prime} + 2 r b^{\prime} + 2a & & {}
\nonumber \\
{} + 3r^2 \eta \left( -r b^{\prime} + a \right) & = & 0
\label{eq:rr}
\end{eqnarray}
Before going further, we can check that the Schwarzschild solution (SS) is a
possible solution of (\ref{eq:rr}) and the trace equation,
(\ref{eq:trace_2}). SS is characterized by $A(r) = 1/B(r) = 1 + \beta/r$,
i.e. $a(r) = -b(r) = \beta/r$ \footnote{A few years ago this writer
speculated \cite{prp15} that a second solution of the W-E equation might
also satisfy these Schwarzschild conditions. The present paper suggests this
idea is mistaken.}. Substituting these expressions into
(\ref{eq:rr}) and (\ref{eq:trace_2}), we can verify that the equations are
satisfied.

In the limit $\alpha_g \rightarrow \infty$ ($\eta \rightarrow 0$), the Weyl
tensor is everywhere dominant. The trace equations are irrelevant, and
(\ref{eq:rr}) admits the solution
\begin{equation}
a(r) = -b(r) = \gamma r\,,
\end{equation}
the Mannheim linear potential.

These are not, however, the only possibilities, and we will now derive a
different form for $a(r)$ and $b(r)$. The solution we will obtain does not,
of course, guarantee that a corresponding solution exists for the full
nonlinear W-E equations. But it does provide a limiting form, for weak
fields, of such a solution, if it exists.

We will now transform (\ref{eq:rr}) and the trace equation to get a
second-order equation in $a(r)$ only. Differentiating (\ref{eq:trace_2}):
\begin{eqnarray}
-4a^{\prime} -2ra^{\prime \prime}
+ 4r b^{\prime \prime} + r^2 b^{\prime \prime \prime}
+ 2b^{\prime} & = & {}
\nonumber \\
 \frac{2r \xi \rho(r)}{\eta}
+ \frac{r^2 \xi \rho^{\prime} (r)}{\eta} & & {}
\end{eqnarray}
Combining this with (\ref{eq:rr}) we can eliminate
$b^{\prime \prime \prime} (r)$:
\begin{eqnarray}
-3r^2 a^{\prime \prime} - 4r a^{\prime} + 2a
+ 2r^2 b^{\prime \prime} + 4 r b^{\prime} \hspace{0.6in} & & {}
\nonumber \\
{} + 3r^2 \eta \left( -r b^{\prime} + a \right) = 
\frac{2r^2 \xi \rho(r)}{\eta}
+ \frac{r^3 \xi \rho^{\prime} (r)}{\eta} & & {}
\end{eqnarray}

We can now use (\ref{eq:trace_2}) and (\ref{eq:beqn}) to eliminate all
terms involving $b(r)$, to arrive at
\begin{equation}
-3r^2 a^{\prime \prime} + \left( 6 - 3r^2 \eta \right) a =
3r \frac{\xi}{4 \pi} m_e (r)
+ \frac{r^3 \xi \rho^{\prime} (r)}{\eta}
\label{eq:2ord}
\end{equation}

We will develop the solution of this equation as an integral over the source
density, using a Green function constructed from the related homogeneous
equation
\begin{eqnarray}
-3r^2 a^{\prime \prime} + \left( 6 - 3r^2 \eta \right) a & = & 0
\label{eq:2ordhom}
\end{eqnarray}

At this point we choose $\alpha_g < 0$, and therefore $\eta < 0$, purely
for computational convenience. This will ensure that we deal with modified
Bessel functions, which have a particularly simple form.

(\ref{eq:2ordhom}) can be written
\begin{eqnarray}
a^{\prime \prime} + \left( -\frac{\nu^2 - 1/4}{r^2} - k^2 \right) a
& = & 0
\label{eq:2ordhom2}
\end{eqnarray}
with $\nu = 3/2$ and $k^2 = -\eta > 0$. Solutions are (see \cite{absteg}
9.1.49) $a(r) = r^{1/2} {\mathcal L}_{3/2} (kr)$, where ${\mathcal L}_{\nu}$
stands for $I_{\nu}$ or $K_{\nu}$.

Because in this paper we are looking for solutions analogous to Mannheim's
linear potential, we will assume, tentatively, that the length $r_0 = 1/k$
is of galactic scale, intermediate between the scale of the Solar System
and truly cosmological scales.

From (\ref{eq:2ordhom2}) we can derive a Green function by standard methods
\cite{arfken2}. If the function is defined on the range $0$ to $\infty$,
with $u$ the source point and $t$ the field point:
\renewcommand{\arraystretch}{1.5}
\begin{eqnarray}
G(u,t) & = & \left\{ \begin{array}{ll}
I_{3/2} (u) K_{3/2} (t), \hspace{0.25in} 0 \le u < t, \\
K_{3/2} (u) I_{3/2} (t), \hspace{0.25in} t < u \le \infty.
\end{array} \right.
\end{eqnarray}
\renewcommand{\arraystretch}{1.0}
so that $a(r)$ can be written as an integral over the source
\begin{eqnarray}
a(r) & = & r^{1/2} \int_{0}^{\infty} G(kr,kt)
\left( \frac{1}{3 t^{1/2}} \right)
\nonumber \\
  & & \times \left[ \frac{3 \xi}{4 \pi} m_e (t)
+ \frac{t^2 \xi \rho^{\prime} (t)}{\eta } \right]\,dt
\label{eq:dfe_green_a}
\end{eqnarray}

It is convenient to define two new functions:
\begin{eqnarray}
\ovl{K}_{3/2} (z) & \equiv &
\frac{K_{3/2} (z) }{\displaystyle z^{1/2} } e^z
 = \sqrt{\frac{\pi }{2}} 
\left(\frac{1}{z} + \frac{1}{z^2 } \right)
\label{eq:kbar}
\\
\ovl{I}_{3/2} (z) & \equiv &
\frac{I_{3/2} (z) }{\displaystyle z^{1/2} } e^{-z}
\nonumber \\
  & = & \sqrt{\frac{2}{\pi}} e^{-z}
\left( -\frac{\sinh z}{z^2 } + \frac{\cosh z}{z} \right)
\nonumber \\
  & & \hspace{-0.5in} = \sqrt{\frac{1}{2 \pi}} \left[
\left(-\frac{1}{z^2 } + \frac{1}{z} \right)
+ e^{-2z} \left(\frac{1}{z^2 } + \frac{1}{z} \right) \right]
\label{eq:ibar}
\end{eqnarray}

In terms of these functions, (\ref{eq:dfe_green_a}) reads
\begin{eqnarray}
a(r)  & = & a_< (r) + a_> (r),\hspace{0.25in}{\textrm where}
\nonumber \\
a_< (r) & = & kr \ovl{K}_{3/2} (kr)
\int_{0}^{r} \ovl{I}_{3/2} (kt) e^{k(t-r)} \left( \frac{1}{3 } \right)
\nonumber \\
  & & \times \left[ \frac{3 \xi}{4 \pi} m_e (t)
+ \frac{t^2 \xi \rho^{\prime} (t)}{\eta } \right]\,dt
\label{eq:dfe_green_a2}
\\
a_> (r) & = & kr \ovl{I}_{3/2} (kr)
\int_{r}^{\infty} \ovl{K}_{3/2} (kt) e^{k(r-t)} \left( \frac{1}{3 }
\right)
\nonumber \\
  & & \times \left[ \frac{3 \xi}{4 \pi} m_e (t)
+ \frac{t^2 \xi \rho^{\prime} (t)}{\eta } \right]\,dt
\label{eq:dfe_green_a3}
\end{eqnarray}

\section{\label{sec:sun}The gravitational field of the Sun}

The most immediate application of our formulae is to obtain
the analog of the Schwarzschild solution, i.e. the metric functions in the
vacuum surrounding a compact massive object such as the Sun. Our
second-order equation, (\ref{eq:2ord}), reduces in this case to
\begin{eqnarray}
-3r^2 a^{\prime \prime} + \left( 6 - 3r^2 \eta \right) a & = &
3r \frac{\xi}{4 \pi} m_e (r)
\label{eq:2ordvac}
\end{eqnarray}

The general solution of this equation is a Particular
Integral (PI) plus a Complementary Function (CF). A suitable PI is
\begin{equation}
a(r) = - \frac{\xi m_{\odot}}{4 \pi \eta r} = \frac{2 G_0 m_{\odot}}{r}
\hspace{0.25in}{\textrm PI\;only}
\label{eq:PIlarge}
\end{equation}
while the CF is proportional to $ r^{1/2} K_{3/2} (kr)$
(contributions from $I_{3/2}$ can be ruled out because they contain a rising
exponential). So we can write
\begin{eqnarray}
a(r) & = & -\frac{\xi m_{\odot}}{4 \pi \eta r }
+ {\mathcal C} \left( 1 + \frac{1}{kr} \right) e^{-kr}\,,
\label{eq:Cdef}
\end{eqnarray}
where we will use our Green function to determine the constant
${\mathcal C}$. For this we need only study the most singular terms for
small $r$.

Let us first consider the limit of large $r$. In the vacuum outside the
source, $m_e (r) = m_{\odot}$ and $\rho^{\prime} = 0$. The exponentials in
the integrals in (\ref{eq:dfe_green_a2}) and (\ref{eq:dfe_green_a3}) will
be sharply peaked around $u = t$. In (\ref{eq:kbar}) and (\ref{eq:ibar})
we will keep only the leading terms, i.e. we neglect $1/z^2$ in comparison
to $1/z$, and omit terms in $\exp(-2z)$ altogether.
\begin{eqnarray}
a_< (r) & = & kr \sqrt{\frac{\pi}{2}} \left(\frac{1}{kr} \right)
\int_{0}^{r} \sqrt{\frac{1}{2\pi}} \left(\frac{1}{kt} \right)
e^{k(t-r)}
\nonumber \\
  & & \times \left[\frac{\xi m_{\odot}}{4 \pi}\right]\,dt
\nonumber \\
  & \approx & \left(\frac{1}{2kr} \right) \left[\frac{\xi m_{\odot}}{4 \pi}
\right] \int_{0}^{r} e^{k(t-r)}\,dt
\nonumber \\
  & = & \left(\frac{1}{2k^2 r} \right) \left[\frac{\xi m_{\odot}}{4 \pi}
\right]
\nonumber \\
  & = & -\left(\frac{1}{2 r} \right) \left[\frac{\xi m_{\odot}}{4 \pi \eta}
\right]
\end{eqnarray}

$a_> (r)$ will contribute an identical amount, so
\begin{equation}
a(r) = -\left(\frac{1}{ r} \right) \frac{\xi m_{\odot}}{4 \pi \eta}
= \frac{2 G_0 m_{\odot}}{r}
\label{eq:alarger}
\end{equation}
in the limit of large $r$. As expected, this is just the PI we obtained
earlier.

For the limit of small $r$ we will consider a point in the Solar System,
outside the source but with $r \ll 1/k$. The Green function equations
(\ref{eq:dfe_green_a2}) and (\ref{eq:dfe_green_a3}) can be divided into four
integrals:

\begin{eqnarray}
a_< (r) & = & kr \ovl{K}_{3/2} (kr) \left[ {\mathcal H}_1 + {\mathcal H}_2
\right], \hspace{0.25in}{\textrm where}
\nonumber \\
{\mathcal H}_1 & = & 
\int_{0}^{r} \ovl{I}_{3/2} (kt) e^{k(t-r)}
\left[ \frac{ \xi}{4 \pi} m_e (t) \right]\,dt
\nonumber \\
{\mathcal H}_2 & = & 
\int_{0}^{r} \ovl{I}_{3/2} (kt) e^{k(t-r)} 
\left[ \frac{t^2 \xi \rho^{\prime} (t)}{3 \eta } \right]\,dt
\end{eqnarray}
and
\begin{eqnarray}
a_> (r) & = & kr \ovl{I}_{3/2} (kr) \left[ {\mathcal H}_3 + {\mathcal H}_4
\right], \hspace{0.25in} {\textrm where}
\nonumber \\
{\mathcal H}_3 & = & 
\int_{r}^{\infty} \ovl{K}_{3/2} (kt) e^{k(r-t)} 
\left[ \frac{ \xi}{4 \pi} m_e (t) \right]\,dt
\nonumber \\
{\mathcal H}_4 & = & 
\int_{r}^{\infty} \ovl{K}_{3/2} (kt) e^{k(r-t)} 
\left[ \frac{t^2 \xi \rho^{\prime} (t)}{3 \eta } \right]\,dt
\end{eqnarray}
The last of these, ${\mathcal H}_4 $, is clearly zero. The first,
${\mathcal H}_1 $, can be simplified by writing $m_e (r) = m_{\odot}$
throughout the range of integration, even inside the Sun; the error incurred
this way is smaller by a factor of $k^2 r_{s}^{2} $ than the dominant terms,
where $r_s $ is the radius of the Sun. We then get:
\begin{eqnarray}
{\mathcal H}_1 & = & \left[\frac{\xi m_{\odot}}{4 \pi } \right] e^{-kr}
\int_{0}^{r} \ovl{I}_{3/2} (kt) e^{kt} \,dt
\nonumber \\
  & = & \frac{\xi m_{\odot} e^{-kr}}{4 \pi }
\sqrt{\frac{2}{\pi}}
\int_{0}^{r} \left(-\frac{\sinh (kt)}{k^2 t^2 } + \frac{\cosh (kt)}{kt}
\right)\,dt
\nonumber \\
  & = & \frac{\xi m_{\odot} e^{-kr}}{4 \pi k}
\sqrt{\frac{2}{\pi}} \left[\frac{\sinh (kr)}{kr} - 1 \right]
\end{eqnarray}
It will be convenient in what follows to divide ${\mathcal H}_1 $ into two
pieces:
\begin{eqnarray}
{\mathcal H}_1 & = & {\mathcal H}_5 + {\mathcal H}_6, \hspace{0.25in}
{\textrm where}
\nonumber \\
{\mathcal H}_5 & = & \frac{\xi m_{\odot} e^{-kr}}{4 \pi k}
\sqrt{\frac{2}{\pi}} \left[\frac{\sinh (kr)}{kr} \right]
\nonumber \\
{\mathcal H}_6 & = & -\frac{\xi m_{\odot} e^{-kr}}{4 \pi k}
\sqrt{\frac{2}{\pi}}
\end{eqnarray}

For ${\mathcal H}_2 $ we get the leading terms by setting
$\ovl{I}_{3/2} (kt)$ equal to its limiting value for small
$kt$,r$\sqrt{1/(2 \pi)} (2kt/3)$:
\begin{eqnarray}
{\mathcal H}_2 & = & \sqrt{\frac{1}{2\pi}} \frac{\xi e^{-kr}}{3 \eta} 
\int_{0}^{r} \left(\frac{2kt}{3} \right) t^2 \rho^{\prime} \,dt
\nonumber \\
  & = & \sqrt{\frac{1}{2\pi}} \frac{2 \xi k e^{-kr}}{9 \eta} 
\left[\left. t^3 \rho (t) \right|_{0}^{r}
- \int_{0}^{r} 3t^2 \rho (t) \,dt \right]
\nonumber \\
  & = & \sqrt{\frac{1}{2\pi}} \frac{\xi m_{\odot} e^{-kr}}{6 \pi k} 
\end{eqnarray}
Finally, ${\mathcal H}_3 $ evaluates to:
\begin{eqnarray}
{\mathcal H}_3 & = & \frac{\xi m_{\odot} e^{kr}}{4 \pi}
\sqrt{\frac{\pi}{2}} \int_{r}^{\infty} 
\left(\frac{1}{kt} + \frac{1}{k^2 t^2 } \right) e^{-kt}\,dt
\nonumber \\
  & = & \frac{\xi m_{\odot} }{4 \pi k^2 r}
\sqrt{\frac{\pi}{2}}
\end{eqnarray}

We get the PI of $a(r)$ from the ${\mathcal H}_5 $ and ${\mathcal H}_3 $
terms:
\begin{eqnarray}
{\textrm PI} & = & kr \ovl{K}_{3/2} (kr) 
\frac{\xi m_{\odot} e^{-kr}}{4 \pi k}
\sqrt{\frac{2}{\pi}} \left[\frac{\sinh (kr)}{kr} \right]
\nonumber \\
  & & {} + kr \ovl{I}_{3/2} (kr) 
\frac{\xi m_{\odot} }{4 \pi k^2 r} \sqrt{\frac{\pi}{2}}
\end{eqnarray}
Using the definitions in (\ref{eq:kbar}) and (\ref{eq:ibar}) we can reduce
this to
\begin{eqnarray}
{\textrm PI} & = & \frac{\xi m_{\odot}}{4 \pi k^2 r} =
-\frac{\xi m_{\odot}}{4 \pi \eta r}
\label{eq:PIsmall}
\end{eqnarray}
in agreement with (\ref{eq:PIlarge}) and (\ref{eq:alarger}), as expected.

The CF is formed from the ${\mathcal H}_2 $ and ${\mathcal H}_6 $ terms:
\begin{eqnarray}
{\textrm CF} & = &  kr \ovl{K}_{3/2} (kr) \frac{\xi m_{\odot} e^{-kr}}{\pi k}
\left( \sqrt{\frac{2}{\pi}} \frac{1}{12}
- \sqrt{\frac{2}{\pi}} \frac{1}{4} \right)
\nonumber \\
  & = &  - \frac{\xi m_{\odot}}{6 \pi k }
\left( 1 + \frac{1}{kr} \right) e^{-kr}
\end{eqnarray}
so altogether
\begin{eqnarray}
a(r) & = & \frac{\xi m_{\odot}}{4 \pi k^2 r} 
- \frac{\xi m_{\odot}}{6 \pi k }
\left( 1 + \frac{1}{kr} \right) e^{-kr}
\label{eq:araw}
\end{eqnarray}
and ${\mathcal C} = -\xi m_{\odot}/(6 \pi k)$, from (\ref{eq:Cdef}).

We can now get $b(r)$ by integrating (\ref{eq:beqn}):
\begin{eqnarray}
b^{\prime} (r) & = & \frac{2a(r)}{r} - \frac{\xi m_{\odot}}{4 \pi k^2 r^2 }
\nonumber \\
  & = & \frac{\xi m_{\odot}}{4 \pi k^2 r^2 }
- \frac{\xi m_{\odot}}{3 \pi } \left( \frac{1}{kr} + \frac{1}{k^2 r^2 }
\right) e^{-kr}
\nonumber \\
b(r) & = & -\frac{\xi m_{\odot}}{4 \pi k^2 r}
+ \frac{\xi m_{\odot}}{3 \pi k} \left( \frac{e^{-kr}}{kr} \right)
\label{eq:braw}
\end{eqnarray}

\begin{figure}[ht]
\centering
\includegraphics[scale=0.27,angle=-90]{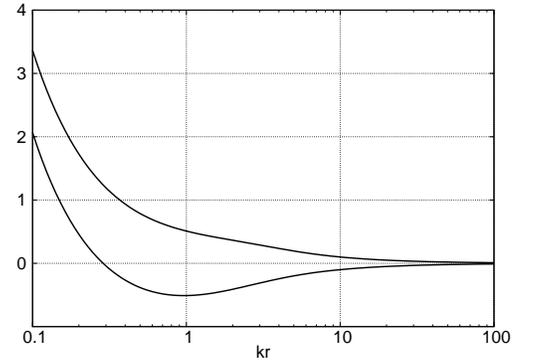}
\caption{\label{fig:abhat} Upper (lower) curve: $a$ ($b$)
as a function of $kr$. The vertical scale is arbitrary. }
\end{figure}

As they stand, these expressions for $a(r)$ and $b(r)$ are unacceptable,
because the leading terms, for small $r$, give
\begin{equation}
a(r) = b(r) = \frac{\xi m_{\odot}}{12 \pi k^2 r}
\end{equation}
whereas observations tell us that in the Solar System we have a
solution very close to Schwarzschild form, for which $a(r) = -b(r)$.
Fortunately a possible remedy is close to hand, as described in the next
section. Before discussing that, however, we should check our solution for
signs of the Mannheim linear potential, which should be evident in the
limit $k \rightarrow 0$. In taking this limit, we keep $\xi/k^2$ fixed at
its value of $8\pi G_0 $. Then (\ref{eq:braw}) gives
\begin {eqnarray}
b(r) & = & \frac{G_0 m_{\odot}}{6r} \left[ -3
+ 4\left( 1 - kr + \frac{k^2 r^2 }{2}  + \cdots \right) \right]
\nonumber \\
  & = & \frac{G_0 m_{\odot}}{6r}
- \frac{G_0 m_{\odot} k}{6}
+ \frac{G_0 m_{\odot} k^2 r}{12} + \cdots
\label{eq:bexpand}
\end{eqnarray}
The Mannheim linear potential has made its appearance, as part of an
approximation to a falling exponential.

\section{\label{sec:get}Getting to a Schwarzschild solution}

Flanagan \cite{flanagan} has pointed out that
the effective metric tensor cannot be $g_{\mu \nu}$, which is not
conformally invariant, but must be
\begin{eqnarray}
\hat{g}_{\mu \nu} & = & F^2 (r) g_{\mu \nu}
\end{eqnarray}
where $F(r)$ is some scalar field of conformal weight $-1$. We will
follow Flanagan in identifying $F(r)$ with $S(r)/m_0 $ where $S(r)$ is
Mannheim's scalar field, and $m_0 $ is some convenient scale of mass,
which we take to be the numerical value of $S_0 $.

After the SBT, $S(r)$ has the form $S_0 [1 + s(r) ]$, where $s(r)$
represents oscillations about the minimum of the potential. We will assume
an equation of motion for $s(r)$:
\begin{equation}
s^{;\mu}_{\;\; ;\mu} - k^2 s = -4\pi {\mathcal D} \rho (r)
\end{equation}
This has the static, point-source solution outside the Sun:
\begin{equation}
s(r) = {\mathcal D} m_{\odot} \frac{e^{-kr}}{r}
\end{equation}

Flanagan's field has the form
\begin{equation}
F(r) = 1 + {\mathcal D} m_{\odot} \frac{e^{-kr}}{r}
\end{equation}

For $\hat{g}_{\mu \nu}$ to approximate a metric of Schwarzschild type we
must have
\begin{eqnarray}
F(r)A(r) = \frac{1}{F(r)B(r)}
\end{eqnarray}
Expnding to first order, we get
\begin{eqnarray}
s(r) & = & -\frac{a(x) + b(x)}{4}
\nonumber \\
{\mathcal D} m_{\odot} \frac{e^{-kr}}{r}  & \approx &
-\left(\frac{\xi m_{\odot}}{24 \pi k^2 r} \right) e^{-kr}
\end{eqnarray}
Since $\xi/k^2 = 8 \pi G_0 $, we have ${\mathcal D} = -G_0/3$.

For our new metric functions, $\hat{a} (r)$ and $\hat{b} (r)$, we get
\begin{eqnarray}
\hat{a}(r) & = & a(r) + 2 s(r)
\nonumber \\
  & = & \left[\frac{\xi m_{\odot}}{4 \pi } \right]
\left( \frac{1}{k^2 r} \right)
\nonumber \\
  & & - \left[\frac{\xi m_{\odot}}{6 \pi } \right]
\left( \frac{1}{k} \right) \left( 1 + \frac{3}{2kr} \right) e^{-kr}
\label{eq:ahat}
\\
\hat{b}(r) & = & b(r) + 2 s(r)
\nonumber \\
  & = & 
- \left[ \frac{\xi m_{\odot}}{4 \pi} \right]
\left(\frac{1}{k^2 r} \right)
\nonumber \\
  & & {} + \left[ \frac{\xi m_{\odot}}{6 \pi} \right]
\left(\frac{1}{k}\right)\left(\frac{3e^{-kr}}{2kr} \right)
\label{eq:bhat}
\end{eqnarray}

\begin{figure}[ht]
\centering
\includegraphics[scale=0.27,angle=-90]{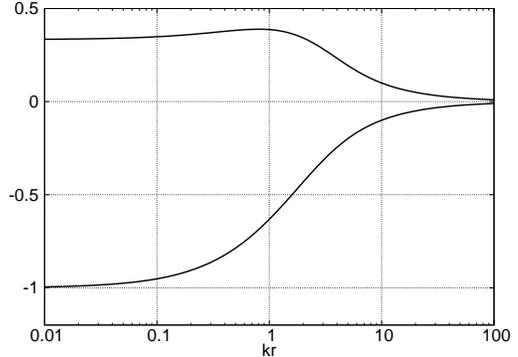}
\caption{\label{fig:abhat} Upper (lower) curve: $\hat{a}$ ($\hat{b}$)
as a function of $kr$. The vertical scale is arbitrary. }
\end{figure}

The surprise here is that the coefficients of the $1/r$ terms for both
$\hat{a}$ and $\hat{b}$ turn out to be zero, so that in getting to a
Schwarzschild form we are obliged to eliminate the normal gravitational
field altogether. We conclude that a solution of Mannheim type (a
Schwarzschild potential plus a linear potential, or an approximation to it)
cannot exist for the W-E equation.

\section{\label{sec:newapp}Can the W-E equation represent reality?}

We have shown that a solution of Mannheim type cannot exist for the W-E
equation. This does not mean, however, that the equation is useless. We
should simply discard our initial assumption, that $r_0 = 1/k$ is of
galactic scale. Indeed, it would be surprising if a SBT resulted in so large
a value of $r_0 $. More likely would seem to be a value of order
$1\,{\textrm fm}$ or less. In this case the Einstein equations would be adequate
at all scales accessible to experiment.


The W-E equation could still have important theoretical applications,
however, because at the highest energies we expect the SBT to be reversed,
so that we recover the original conformal form in which all coupling
constants are dimensionless. The theory is then potentially renormalizable.



 
\end{document}